\documentclass[twocolumn,showpacs,preprintnumbers,amsmath,amssymb,prl,epsf]{revtex4}
\usepackage{graphicx}
\usepackage{color}
\newcommand{\be}{\begin{equation}}
\newcommand{\ee}{\end{equation}}
\newcommand{\bea}{\vspace{0.25cm}\begin{eqnarray}}
\newcommand{\eea}{\end{eqnarray}}
\begin{document}
\title{Characterization of Spectral Entanglement of Spontaneous Parametric-Down Conversion Biphotons }
\author{G. Brida$^1$, V. Caricato$^{1,\,2}$, M. Genovese$^1$, M. Gramegna$^1$}
\affiliation{$^1$I.N.RI.M.- Istituto Nazionale di Ricerca Metrologica, Turin,Italy \\
 $^2$Department of Physics, Politecnico of Turin, Turin,Italy}

\author{M.~V.~Fedorov}
\affiliation{A.M.Prokhorov General Physics Institute, Russian Academy of Science, Moscow, Russia}

\author{S.~P.~Kulik}
\affiliation{Faculty of Physics, M.V.Lomonosov Moscow State University, Moscow, Russia}\vskip 24pt

\date{\today}
\begin{abstract}
We verified operational approach based on direct measurement of
entanglement degree for bipartite systems. In particular spectral
distributions of single counts and coincidence for pure biphoton
states generated by train of short pump pulses have been measured
and entanglement quantifier calculated. The approach gives upper
bound of entanglement stored in total biphoton states,
which can reach extremely high value up to $10^{4}-10^{5}$.

\end{abstract}
\pacs{42.65.Lm, 03.65.Ud, 03.67.Mn}\maketitle

Recently entanglement became from an argument of debate on
foundations on quantum mechanics a resource for developing quantum
technologies \cite{mg} (as quantum communication, q-calculus,
q-imaging, q-metrology, etc.). In this sense a precise and
easy-implementable method for characterizing entanglement properties
is a fundamental tool. There are several measures that quantify
entanglement of quantum bipartite states both in discrete and
continues variables \cite{en}. The most popular ones are Schmidt
rank, entropy, concurrence, etc. From a theoretical point of view
all these quantifiers can be easily evaluated from the density
matrix describing the given state, thus  there is no problem to find
the meaning of a chosen entanglement measure for  a certain density
matrix. From an experimental side and following the same paradigm it
would be necessary to perform quantum tomography procedure (either
complete or reduced) over the state \cite{tom} under study
and then to extract the entanglement measure from reconstructed
density matrix. However this scheme is not optimal in the case of
high-dimensional systems when number of measurements for complete
reconstruction of the density matrix grows quadratically with
dimension of Hilbert space. An alternative is exploiting particular
links between entanglement and measurable characteristics of the
state, just measuring some auxiliary parameters of the state. In
particular the parameter $R_{q}$ (Fedorov's ratio \cite{Fedorov06}),
defined as the ratio of the single-particle and coincidence
distributions widths in $q$-space, can be rather easily measured in
contrast to all other entanglement quantifiers.  $R$ can be
understood qualitatively from entropy approach to entanglement: high
entanglement leads to the better knowledge about composite bipartite
system (narrower coincidence distribution) and the worse knowledge
about individual subsystem(s) (wider single-particle distribution).
Both clear physical meaning and operationability return $R$
to an extremely useful tool for entanglement control. It has been
proved \cite{Fedorov06} that for double-Gaussian bipartite states
the parameter $R$ coincides exactly with the Schmidt number $K$.
Moreover their values remain quite close even for special classes of
non-double-Gaussian wave functions like those describing Spontaneous
Parametric Down-Conversion (SPDC). This approach  can be used both
for continues and discrete variables, so the parameter R has rather
universal sense: earlier it has been successfully applied for
exposure strong entanglement anisotropy in spatial distributions of
biphotons \cite{angPRAL}.

The present paper is devoted to verification of the operational
entanglement quantifier $R$ for two-photon states entangled in
frequency domain as an efficient alternative to other quantifiers
(for example in specific experiments visibility of
interference pattern can provide some knowledge about entanglement
\cite{U'Ren}). Indeed such states belong to multi-dimensional
Hilbert space and can posses extremely high entanglement degree (up
to several hundreds) that makes its very perspective objects for
quantum information and quantum communication.

Let us remind first the main features of the biphoton spectral wave
function for SPDC with the type-I degenerate collinear phase
matching and with a pulsed pump. The main general expression has the
form \cite{Spectral,Rubin}
\begin{eqnarray}
 \nonumber
 \Psi(\nu_1,\,\nu_2)\propto \exp\left(-\frac{(\nu_1+\nu_2)^2\tau^2}{8\ln
 2}\right)\\
 \times
 {\rm  sinc}\left\{\frac{L}{2c}\left[A(\nu_1+\nu_2)-B\frac{(\nu_1-\nu_2)^2}{\omega_p}\right]
 \right\},\label{WF}
\end{eqnarray}
where  $\tau$ is the pump-pulse duration, $L$ is the length of the
crystal, $\nu_1$ and $\nu_2$ are deviations of frequencies of the
signal and idler photons $\omega_{1,\,2}$ from the central
frequencies $\omega_1^{(0)}=\omega_2^{(0)}=\omega_p/2$,
$|\nu_{1,\,2}|\ll\omega_p$, $\omega_p$ is the central frequency of
the pump spectrum, $A$ and $B$ are the temporal walk-off and
dispersion constants
\begin{gather}
 \nonumber
 A=c\left(\left.k_p^\prime(\omega)\right|_{\omega=\omega_p}-
 \left.k_1^\prime(\omega)\right|_{\omega=\omega_p/2}\right)=c\left(\frac{1}{{\rm v}_g^{(p)}}-\frac{1}{{\rm v}_g^{(o)}}\right)\label{A},\\
 \label{AB}
 B=\frac{c}{4}\,\omega_p\left.k_1^{\prime\prime}(\omega)\right|_{\omega=\omega_0/2},
\end{gather}
${\rm v}_g^{(p)}$ and ${\rm v}_g^{(o)}$ are the group velocities of
the pump and ordinary waves, and $k_1$ and $k_p$ are the wave
vectors of signal and pump photons.


The wave function ({\ref{WF}}) can be served for determining coincidence and single-particle biphoton spectra. These spectra are
significantly different in the cases of short an long pump pulses.
The control parameter separating the regions of short and long
pulses is given by \cite{Spectral}
\begin{equation}
 \label{eta}
 \eta=\frac{\Delta\nu_{1\,{\rm sinc}}}{\Delta\nu_{1\,{\rm pump}}}
 \approx 2\frac{c\,\tau}{AL}=\frac{2\tau}{L/{\rm v}_g^{(p)}-L/{\rm v}_g^{(o)}},
\end{equation}
i.e., it is equal to the ratio between the double pump-pulse
duration and the difference of times required for the the pump and
idler/signal photons for traversing all the crystal. Pump pulses are
short if $\eta\ll 1$ and long if $\eta\gg 1$ and, typically,
$\eta\sim 1$ at $\tau\sim 1 {\rm ps}$.

In the most interesting case of short pump pulses the FWHM of the
coincidence and single-particle spectra were found analytically to
be given by \cite{Spectral}
\begin{equation}
 \label{FWHM}
 \Delta\omega_c=\frac{5.56\,c}{AL}, \quad\Delta\omega_s=\sqrt{\frac{2A\ln (2)\,\omega_p}{B\tau}}.
\end{equation}

The degree of entanglement of the state (\ref{WF}) was characterized
by two parameters, the Schmidt number $K$ and the parameter $R$
defined as the ratio of the single- to coincidence spectral widths
of the corresponding photon distributions,
$R_{\omega}=\Delta\omega_s/\Delta\omega_c\approx\Delta\lambda_s/\Delta\lambda_c=R_{\lambda}$ \cite{Spectral}.
\begin{figure}[h]
\centering\includegraphics[width=6cm]{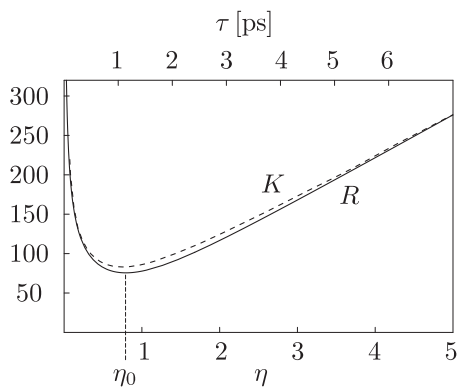}
\caption{{\protect\footnotesize {Quantifiers calculated analytically ($R_{\omega}$ \cite{Spectral}) and numerically ($K$ \cite{Mauerer}) for LiIO$_3$ crystal and the pump wavelength
$\lambda_p=400\,{\rm nm}$}.}}\label{Fig1}
\end{figure}
Both parameters were calculated analytically as function of the
pump-pulse duration $\tau$ and were found to be very close
to each other. Also, for the same wave function (\ref{WF}),
$K(\tau)$ was calculated numerically \cite{Mauerer}, and the result
was found to be in a very good agreement with the analytical
function $R(\tau)$ of Ref. \cite{Spectral}, Fig. \ref{Fig1}. The
degree of spectral entanglement reaches rather high values in the
whole range of pump-pulse durations and especially high in the case
of sufficiently short and long pulses (far from the minimum that
localizes at $\eta\sim 1$ or $\tau\sim 1\,{\rm ps}$).

Evidently the parameter (\ref{eta}) accumulates factors to be rather
easily controlled in experiment, namely length of the crystal and
pump pulse duration. Also the longitudinal walk-off effect can be
evolved via $A$ by choosing a crystal with appropriate
dispersion properties as a source of biphotons.

The experiment(Fig.2) was performed in fs pulsed regime
using a Mode-Locked Titanium-Sapphire laser at a working wavelength
of $\lambda_{IR}=(795.0 \pm 0.1)$ nm with a $\Delta\lambda_{IR}=(5.9
\pm 0.1)$ nm. After doubling in frequency to $\lambda_p=(397.5 \pm
0.2)$ nm and $\Delta \lambda_p=(1.8 \pm 0.1)$ nm, corresponding to a
pulse duration of $\tau= (186 \pm 30)$ fs, the pump beam is
addressed to a LiIO$_{3}$ crystal where type I collinear SPDC is
produced. After eliminating the UV pump, biphotons are split on a
beam-splitter and  fed to two photodetection apparatuses (consisting
of red glass filters and SPAD detectors). In front of each detector
it is placed a monochromator with a variable spectral resolution.
Light is focused by a lens ($f$=20 cm) inside each
monochromator and a couple of lenses with the same focal distance
are placed on the output of the monochromators for focalizing the
output packet spread by all the optics of the apparatus and by the
slits.

\begin{figure}[h]
\includegraphics[width=9cm]{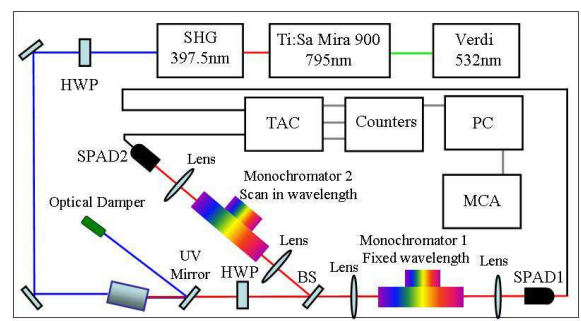}
\caption{Sketch of the experimental set up: a duplicated titanium
sapphire laser beam pumps a LiIO$_3$ crystal producing collinear
SPDC. After a Half Wave Plate (HWP) the biphotons are split on a
beam-splitter (BS) and fed to  SPAD detectors, whose output
feeds Time-to Amplitude Converter (TAC), counters and Multi Channel
Analyzer (MCA).}
\end{figure}

Following to general idea of the $R$-quantifier method all
experiments were performed as a consecution of corresponding
measurements of spectral distributions both in single counts and
coincidences. To measure the coincidence distribution one of the two
monochromators (by convention selecting idler wavelength) is fixed
at the central wavelength of SPDC (795 nm) whilst the other (signal)
scans in a range around this value. In order to check $R$
 dependence on the control parameter $\eta$ the measurements have
been repeated for crystals with different length (but with the same
orientation), namely $L$=10 mm and $L$=5 mm.

First of all we have performed measurements with a 10 mm LiIO$_3$
crystal. To measure the coincidence distribution we have used
monochromators with 0.2 nm resolution. A narrow
peak with $\Delta\lambda_c = (0.29 \pm 0.03)$ nm has been obtained (Fig.3) while scanning the wavelength in signal channel. It is clearly seen that the pump width is 6.2
times larger than the coincidence spectrum.
\begin{figure}[h]
\centering\includegraphics[width=6cm]{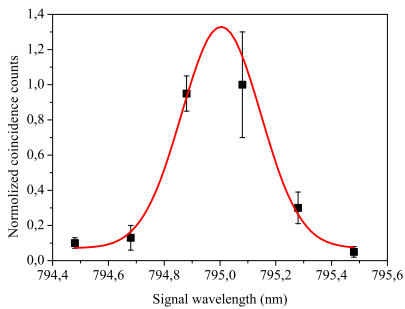}
\caption{{\protect\footnotesize {Normalized coincidence distribution
for 10 mm length LiIO$_3$ sample. Points with error bars correspond
to the measurements while the solid line shows the fitting by a
Gaussian function with FWHM =$(0.29 \pm 0.03)$nm.}
}}\label{Fig3}
\end{figure}

The spectral distribution of single counts was performed using a monochormator in
the transmission arm with a spectral resolution
of 1 nm, finding a single counts spectral width $\Delta \lambda_s = ( 101 \pm 1 )$ nm (Fig.4). The asymmetry of the right wing in the measured spectrum might be caused by falling spectral sensitivity of monochromator in the long-wavelength range.

\begin{figure}[h]
\centering\includegraphics[width=6cm]{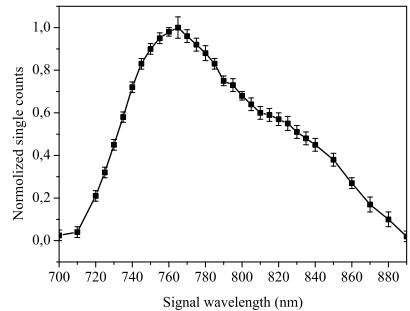}
\caption{{\protect\footnotesize {Normalized single counts distribution for 10 mm length LiIO$_3$ sample}. }}\label{Fig4}
\end{figure}

Thus the distribution of single counts is 56 times larger than that of the pump, corresponding to an
experimental ratio between widths of the two distributions $R_\omega=(349 \pm 43)$. The complete results are collected in the table I.

\begin{table}[!ht]
\caption{Results for 10 mm LiIO$_3$ sample. The pump parameters
$\tau$ and $\Delta\lambda_p$ are measured independently by using a
cross-correlator (with first harmonic) and a spectrometer.}
\begin{tabular}{|c|c|}
 \hline Theory Model&
Experimental results\\
\hline \multicolumn{2}{|c|}{Pump} \\ \hline
\begin{tabular}{c}
$\lambda_p=397.5 nm$ %
\end{tabular}
&
\begin{tabular}{c}
$\lambda_p=397.5 \pm 0.2 nm$ %
\end{tabular}\\ \hline
\begin{tabular}{c}

$\Delta\lambda_p=1.25 nm$ %
\end{tabular}
&
\begin{tabular}{c}
$\Delta\lambda_p=(1.8\pm 0.1) nm$ %
\end{tabular} \\ \hline

\begin{tabular}{c}
$\tau=186 fs$ %
\end{tabular}
&
\begin{tabular}{c}
$\tau=186\pm 30 fs$ %
\end{tabular} \\ \hline

\hline \multicolumn{2}{|c|}{Coincidence Distribution} \\ \hline

\begin{tabular}{c}
$\lambda_c= 795 nm$ %
\end{tabular}
&
\begin{tabular}{c}
$\lambda_c=795.0\pm 0.2 nm$ %
\end{tabular} \\ \hline

\begin{tabular}{c}
$\Delta\lambda_c= 0.32 nm$ %
\end{tabular}
&
\begin{tabular}{c}
$\Delta\lambda_c= (0.29 \pm 0.03)$ %
\end{tabular} \\ \hline

\hline \multicolumn{2}{|c|}{Single Counts Distribution} \\ \hline

\begin{tabular}{c}
$\Delta\lambda_s = 100 nm$ %
\end{tabular}
&
\begin{tabular}{c}
$\Delta\lambda_s= (101\pm 1) nm$ %
\end{tabular} \\ \hline

\hline \multicolumn{2}{|c|}{$R$-Quantifier} \\ \hline

\begin{tabular}{c}
$R_{\omega}\approx R_{\lambda}\approx 316$ %
\end{tabular}
&
\begin{tabular}{c}
$R_{\omega}\approx R_{\lambda}\approx (349 \pm 43)$ %
\end{tabular} \\ \hline

\end{tabular}
\end{table}

Then the same measurement has been performed with a crystal
of 5 mm length obtaining $\Delta\lambda_s=(115 \pm 1) $ nm and
$\Delta\lambda_c=(0.64 \pm 0.06)$ nm, corresponding to $R_\omega
= (179 \pm 18)$. Considering that the pulse duration of the pump did
not change we can observe that in this case the pump is 2.8 times
greater than the coincidence spectrum whereas the single counts
distribution is 64 times larger than the width of the pump. Corresponding results are listed in the table II.

\begin{table}[!ht]
\caption{Results for 5 mm LiIO$_3$ sample }
\begin{tabular}{|c|c|}
 \hline Theory Model&
Experimental results\\
\hline \multicolumn{2}{|c|}{Coincidence Distribution} \\ \hline

\begin{tabular}{c}
$\Delta\lambda_c= 0.63 nm$ %
\end{tabular}
&
\begin{tabular}{c}
$\Delta\lambda_c= (0.64 \pm 0.06)$ %
\end{tabular} \\ \hline

\hline \multicolumn{2}{|c|}{Single Counts Distribution} \\ \hline

\begin{tabular}{c}
$\Delta\lambda_s = 100 nm$ %
\end{tabular}
&
\begin{tabular}{c}
$\Delta\lambda_s= (115\pm 1) nm$ %
\end{tabular} \\ \hline

\hline \multicolumn{2}{|c|}{$R$-Quantifier} \\ \hline

\begin{tabular}{c}
$R_{\omega}\approx R_{\lambda}\approx 158$ %
\end{tabular}
&
\begin{tabular}{c}
$R_{\omega}\approx R_{\lambda}\approx (179 \pm 18)$ %
\end{tabular} \\ \hline

\end{tabular}
\end{table}

Certainly the most comprehensive contribution to the experimental verification of the $R$-quantifier validity in frequency domain would be
 testing dependence $R$ on the control parameter $\eta$ through  all available parameters. Agreement between the measured and predicted
  shapes of the curve shown at Fig.1 would confirm completely the adequacy of the $R$-quantifier approach. Looking at (\ref{eta}, \ref{FWHM})
   it is seen that the "working" parameters might be $\tau$ and $L$.
   However it is difficult to change the pump pulse duration in a wide range with the same laser (from hundreds of $fs$ to dozens of $ps$)
   and pass through the minimum at the curve $R(\eta)$ varying this parameter only. That is why we took into account the fact of linear
   dependence of $R$ on the sample length $L$ through the coincidence distribution width (\ref{FWHM}).
   So doubling the sample length leads to doubling the entanglement degree for the short pump pulse regime of SPDC.
   This fact is clearly illustrated by the obtained results, demonstrating the validity of the conceptual scheme even if, unfortunately,
    subsequent decrease of the sample length reduces  the biphoton flux  too much for measuring widths $\Delta\lambda_s$ and reaching
     the minimum of $R(\eta)$.

 One should mention that in some works (see, e.g., \cite{Mosley08}) the 2D- photon distribution was measured and plotted in the
 $(\omega_1, \omega_2)$ plane. Surprisedly that such 2D-distributions are very useful for making entanglement analysis of biphoton states in context
  of conditional pure state preparation, dispersion broadening \cite{Kim, Walmsley00}, etc.
  This would allow  evaluating $R$-quantifier directly. We would
  like also to mention that a  measurement of spectral width could
  also be obtained by transforming spectrum in time difference
  through a fiber \cite{fiber}.

Finally the question arises "how much entanglement can be stored in
a given state". We remind that the method of measurement of $R$ in
frequency domain implies that other degrees of freedom of biphotons
should be fixed. First off all we mean spatial distribution of
photon pairs emitted by crystal, which must be illuminated by
selecting particular narrow-angular segments with small pinholes.
Vice versa, measurement of $R$-quantifier in spatial variables
($R_{angle}$) implies fixing frequencies of both photons, which is
usually performed with narrow-band filters \cite{angPRAL}. Therefore
the total entanglement potentially stored in the SPDC state can be
conventionally accepted as a product $R_{angle}*R_{\omega}$. On the
one hand this is a good convention since quantifier $R$ has
the physical meaning of effective Hilbert space, so the
total dimension of biphoton states in joint
polarization-frequency-spatial space would be $R_{tot} \approx
2*R_{angle}*R_{\omega}$. However we suggest to use this just as an
upper bound of the stored entanglement taking into account the
existing links between frequency and spatial degrees of freedom
caused by phase matching. Speaking in terms of numbers this value
can reach really huge amount $R_{tot} \approx 10^4-10^5$
\cite{insert1}!

In conclusion we have verified the approach  of Ref.
\cite{Fedorov06} addressed to estimate the entanglement degree in
frequency domain for pure biphoton states generated by short pump
pulses. An approach that can find widespread application in quantum
technologies exploiting entanglement of biphotons. The
generalization to the mixed two-photon states will be discussed
elsewhere.

This work has been supported in part by MIUR (PRIN 2007FYETBY),
Regione Piemonte (E14),  "San Paolo foundation", NATO (CBP.NR.NRCL
983251) and RFBR (08-02-12091).


\begin{references}
\bibitem{mg} M. Genovese, Phys. Rep. 413, 3197 {\bf (2005)}.
\bibitem{en} I. Bengtsson and K. Zyczkowski, Geometry of Quantum
States, Cambridge Univ. press {\bf (2006)}.
\bibitem{tom} A.I. Lvovsky et al, Phys. Rev. Lett. 87 (2001) 050402;
A. Zavatta et al., Phys. Rev. A 70, 053821 (2004); A. Allevi et al.,
arXiv:0903.0104.

\bibitem{Fedorov06} M.V. Fedorov, M.A. Efremov, P.A. Volkov, and J.H. Eberly, J. Phys. B: At. Mol. Opt. Phys., \textbf{9}, S467 (2006).
\bibitem{angPRAL}M.V. Fedorov, M.A. Efremov, P.A. Volkov, E.V. Moreva, S.S. Straupe, S.P. Kulik, Phys. Rev. Lett, \textbf{99}, 063901 (2007); Phys. Rev. A, \textbf{77}, 032336 (2008).
\bibitem{U'Ren} Y. Kim et al., Phys. Rev. A 62 (2000) 043820; Y. Jeronimo-Moreno and A. B. U'Ren, Phys. Rev. A \textbf{79}, 033839 (2009).
\bibitem{Spectral} Yu.M. Mikhailova, P.A. Volkov, and M.V.Fedorov, Phys. Rev. A {\bf 78}, 062327 (2008).
\bibitem{Rubin} T.E. Keller and M.H. Rubin, Phys. Rev. A \textbf{56}, 1534 (1997).
\bibitem{Mauerer} W. Mauerer and C. Silberhorn, (to be published in QCMC'08 Proceedings AIP, (2008).
\bibitem{Mosley08} P.J. Mosley, J.S. Lundeen, B.J. Smith, P.Wasylczyk, A.B. U'Ren, Ch. Silberhorn, and I.A. Walmsley, Phys. Rev. Lett.,
 \textbf{100}, 133601 (2008); Alejandra Valencia, Alessandro Ceré, Xiaojuan Shi, Gabriel Molina-Terriza, and Juan P.
 Torres, Phys. Rev. Lett. 99, 243601 (2007); M. Avenhaus et al., arXiv 0810.0998.
\bibitem{fiber} G. Brida, M. Chekhova, M. Genovese, M.Gramegna, L. Krivitsky. Phys.
Rev. Lett. 96, 143601 (2006); G. Brida, M. Genovese, L. A.Krivitsky,
M. V. Chekhova, Phys. Rev. A 75, 015801 (2007);  M. Avenhaus, A.
Eckstein, P. J. Mosley, C. Silberhorn,   arXiv:0902.3364 (2008).
\bibitem{Kim} S. Y. Baek, O. Kwon, and Y. H.Kim. Phys. Rev. A \textbf{77}, 013829 (2008); S. Y. Baek, O. Kwon, and Y. H.Kim. Phys. Rev. A \textbf{78}, 013816 (2008).
\bibitem{Walmsley00}R. Erdmann, D. Branning, W. Grice, and I. A. Walmsley. Phys. Rev. A \textbf{62}, 053810 (2000).
\bibitem{insert1} Previous research \cite{angPRAL} has demonstrated that extremely high entanglement degree can be accumulated in spatial variables of biphoton states ($R_{angle}\approx 300$).


\end{references}
\end{document}